\documentclass[10pt,conference]{IEEEtran}
\IEEEoverridecommandlockouts
\usepackage{cite}
\usepackage{amsmath,amssymb,amsfonts}
\usepackage{algorithmic}
\usepackage{graphicx}
\usepackage{textcomp}
\usepackage{hyperref}
\usepackage{flushend}
\usepackage{comment}
\usepackage{mdframed}

\def\BibTeX{{\rm B\kern-.05em{\sc i\kern-.025em b}\kern-.08em
    T\kern-.1667em\lower.7ex\hbox{E}\kern-.125emX}}

\begin{document}

\title{FlakeRanker: Automated Identification and Prioritization of Flaky Job Failure Categories}

\author{
\IEEEauthorblockN{Henri A\"idasso}
\IEEEauthorblockA{
\textit{École de technologie supérieure}\\
Montreal, Canada \\
henri.aidasso.1@ens.etsmtl.ca}

}

\maketitle

\begin{abstract}
This document presents the artifact associated with the ICSE SEIP 25 paper titled “On the Diagnosis of Flaky Job Failures: Understanding and Prioritizing Failure Categories”.
The original paper identifies and analyzes 46 distinct categories of flaky job failures that developers encounter, using Recency (R), Frequency (F), and Monetary (M) measures. In addition, it uses an RFM clustering model to identify and prioritize the most wasteful and persistent. The original paper only discusses the rankings and evolution of the top 20 categories in the results. This artifact contains (1) the regex and scripts used to automate the labeling process for RQ1, (2) complete analysis results, including the ranking of all 46 categories by cost in RQ2 and the evolution of these categories over time in RQ3, and (3) the RFM dataset and scripts used to create the RFM clustering model for prioritization in RQ4. In addition, we engineered the labeling tool and the RFM-based prioritization methodology in a command-line interface (CLI) called \textsc{FlakeRanker} to facilitate reuse and repurposing in future studies.
\end{abstract}

\section{Introduction}

A major challenge hindering the benefits of continuous integration and continuous deployment (CI/CD) pipelines is flaky job failures stemming from issues unrelated to the code changes. When faced with such failures, developers often resort to repeatedly rerunning pipeline jobs or investing significant time in diagnosing the root cause, leading to considerable waste of infrastructure resources and reduced productivity. Prior studies on the topic mainly focused on flaky tests, which however represent only a subset of flaky job failures. Other studies \cite{lampel_when_2021, olewicki_towards_2022} developed automated techniques for detecting flaky jobs to minimize wasteful job reruns. Although valuable, simply detecting flaky jobs is insufficient in software practice to address the waste associated with flakiness. At TELUS -- a large telecommunication company -- flaky job failures arise from various factors extending well beyond test flakiness and necessitating targeted diagnosis and fix strategies that can take up to 48 hours before resolution. To effectively address such inefficiencies, we conduct a study to (1) identify the different categories of flaky job failures and (2) prioritize them to focus efforts on the most prevalent, costly, and current ones.

This paper introduces the artifact \cite{aidasso_diagnosis_2025} associated with the ICSE SEIP 25 paper titled “On the Diagnosis of Flaky Job Failures: Understanding and Prioritizing Failure Categories" \cite{aidasso_icse_2025}. The artifact comprises (1) complete analysis figures along with intermediary result datasets to support partial replication, and (2) \textsc{FlakeRanker}, a command-line (CLI) tool to facilitate the reuse of the automated labeling tool and the RFM prioritization method, with a curated dataset of build jobs from the open-source Veloren\footnote{\url{https://gitlab.com/veloren/veloren}} project to demonstrate its functionalities.

\section{Overview of the Artifact}

The artifact is stored in a permanent Figshare repository \cite{aidasso_diagnosis_2025}. It includes two main components separated into directories as follows: \textbf{(1)} Original Jupypter Notebooks used for the study and full results obtained -- in the \texttt{\textbf{study/}} directory, and \textbf{(2)} Source code of the \textsc{FlakeRanker} CLI tool developed using scripts in the notebooks to facilitate reuse --  in the \texttt{\textbf{src/flakeranker/}} directory. 

\subsection{Study's Jupyter Notebooks and Results}
\label{sec:notebooks}

To conduct the study, we used Jupyter Notebooks because they provide interactivity and have facilitated the communication of results with our industrial partner. The notebooks are thoroughly documented to guide readers step-by-step through the methodology for addressing each RQ, with the corresponding code and the resulting outputs. These notebooks along with complete analysis results obtained are made available in the \texttt{\textbf{study/}} directory, organized into subdirectories as follows:

\begin{itemize}
    \item \texttt{\textbf{data\_labeling\_process/}}. Includes Jupyter Notebooks used for identifying flaky job failure categories and developing the labeling tool, as described in Section III of the original paper.

    \item \texttt{\textbf{rfm\_analysis/}}. Includes the Jupyter Notebook used for investigating the frequency (RQ1), monetary cost (RQ2), and temporal evolution (RQ3) of the 46 identified flaky job failure categories.

    \item \texttt{\textbf{rfm\_prioritization/}}. Includes the Jupyter Notebook used to conduct the RFM model clustering and prioritization (RQ4) of the 46 flaky job failure categories.

    \item \texttt{\textbf{results/}}. Includes additional analysis results obtained during the study such as (a) regex patterns developed for the labeling tool in RQ1, (b) full figures\footnote{The original paper only presents the top 20 categories in RQ2 and RQ3} depicting monetary costs in RQ2 and temporal evolution in RQ3 of all the 46 studied failure categories, (c) complete \texttt{.csv} datasets of RFM values and scores computed for each category in RQ4, and (d) clustering model dump and \texttt{.csv} clustering results discussed in RQ4.
\end{itemize}

\begin{figure*}[ht]
    \centering
    \includegraphics[width=\linewidth]{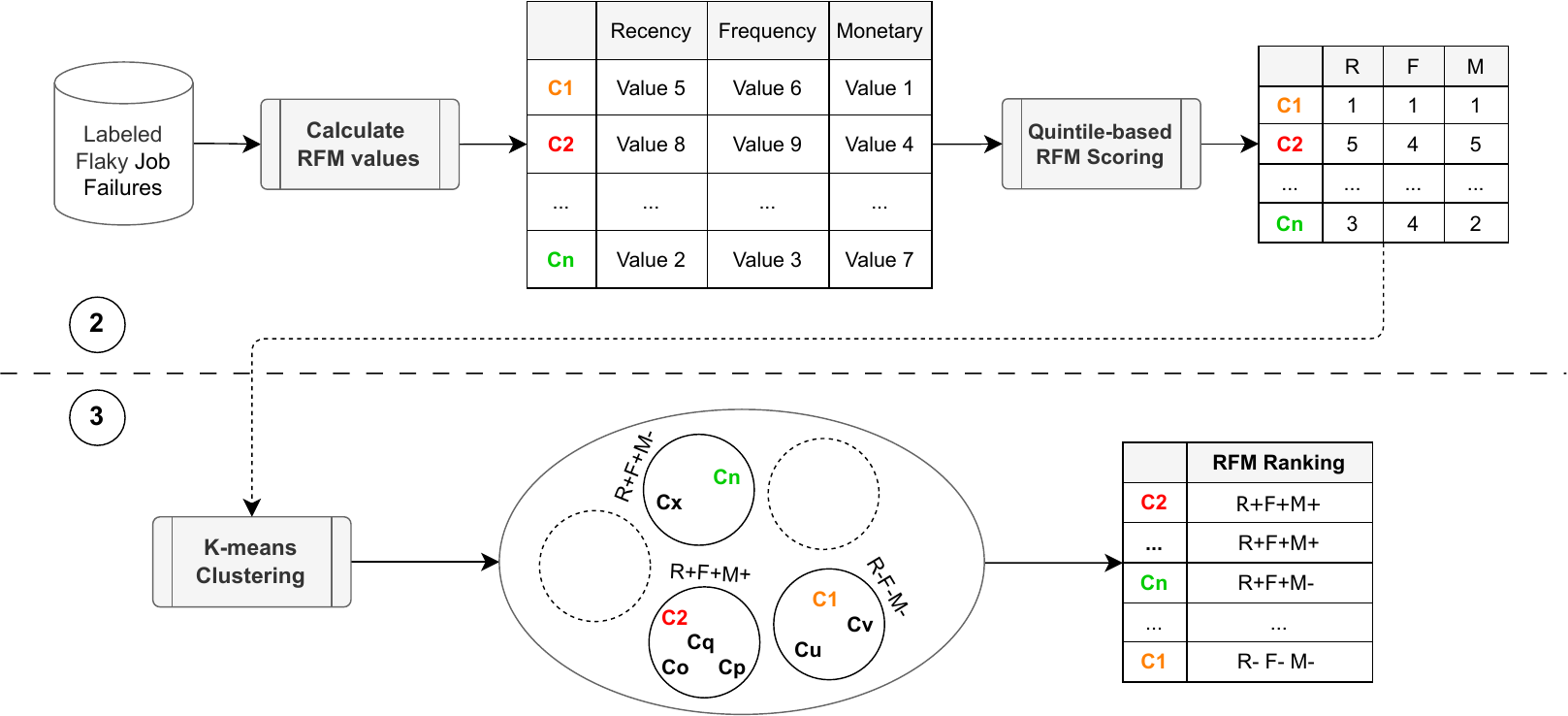}
    \caption{Overview of the FlakeRanker CLI tool, highlighting the \texttt{analyzer} (2) and \texttt{ranker} (3) commands.}
    \label{fig:flakeranker}
\end{figure*}

\subsection{FlakeRanker CLI Tool}

We transformed the notebook scripts into a modular and reusable CLI tool named \textsc{FlakeRanker}. Fig~\ref{fig:flakeranker} illustrates the main implementation steps of the CLI tool. 
While Jupyter Notebooks were convenient for conducting the study due to the reasons outlined in Section~\ref{sec:notebooks}, they are not ideal for reuse, which motivated our decision to develop a corresponding CLI tool. Its source code is made openly available in the \textbf{\texttt{src/flakeranker/}} directory and features three modules (associated with sub-commands), corresponding to the different stages of the study, as described below.

\begin{enumerate}
    \item \textbf{\texttt{labeler/}}. Invoked via the sub-command {\texttt{{label}}}, this module includes the labeling tool scripts for identifying flaky jobs and flaky job failure categories. It takes as input a dataset of jobs (with metadata and execution logs) and outputs a labeled dataset of jobs with the additional columns \textit{flaky} and \textit{category}.

    \item \textbf{\texttt{analyzer/}}. Invoked via the sub-command {\texttt{{analyze}}}, this module includes scripts for calculating each identified category's recency, frequency, and monetary cost measures. As such, it takes as input a labeled dataset of jobs (labeler's output) and outputs an RFM dataset containing the RFM values of each category.

    \item \textbf{\texttt{ranker/}}. Invoked via the sub-command {\texttt{{rank}}}, it includes scripts for calculating RFM scores, fitting the clustering model, and ranking each category based on the RFM patterns. It takes as input an RFM dataset (analyzer's output) and outputs a ranked RFM dataset including the columns \textit{cluster} and \textit{pattern} indicating the cluster and ranking of each category, respectively.
\end{enumerate}

The main command \texttt{{flakeranker} {run}} combines the three aforementioned modules to streamline the labeling and RFM prioritization method as follows:

\begin{mdframed}
\begin{center}
    (input) $=>$ label $=>$ analyze $=>$ rank $=>$ (output)
\end{center}
\end{mdframed}

In addition to the components above, the following documentation is provided in the artifact:

\begin{itemize}
    \item \textit{README.md}: Describes the structure of the entire
    artifact, including the notebooks for addressing each RQ and the locations of additional results. It also includes instructions for building the
    Docker image and executing the prioritization method with \textsc{FlakeRanker} and the example dataset. Finally, it states the badges applicable to this artifact, i.e., the “\textbf{Available}” and “\textbf{Reusable}” badges.

    \item \textit{example/README.md}: Contains detailed documentation of the \textsc{FlakeRanker} commands. In particular, it illustrates and describes the data format required as input and the types of outputs produced by executing each step (i.e. sub-command) of the prioritization approach.
    
    \item \textit{PAPER.pdf}: A copy of our ICSE SEIP 25 accepted paper.
    
    \item \textit{LICENSE}: MIT copyright license
    that grants permission to facilitate the open use and
    distribution of the artifact.
\end{itemize}

\section{Technical Information}

\textbf{Dataset of Build Jobs}. As stated in the original paper, the build jobs dataset from TELUS is not included for confidentiality reasons. Nevertheless, to demonstrate the reuse of \textsc{FlakeRanker}, we have collected and prepared -- in the {\texttt{example/data/veloren.zip}} file -- a dataset of 57,351 build jobs from the Veloren project hosted on GitLab. The zip file also contains a labeled version of the dataset to facilitate the artifact evaluation. In fact, during our tests, the labeling process took about 34 minutes to complete on an Ubuntu 22.04 RAM 16GB Dual Core i7 2.80GHz. Providing the labeled dataset as input bypasses the regex searches involved, allowing for faster execution ($<$ 1min). Hence, we recommend evaluating the CLI tool with the labeled dataset as input and then using the clean dataset afterward if time permits.

\textbf{Technical Requirements}. The artifact is self-contained and requires no additional software dependencies beyond a Docker Engine with a minimum version of 24.0.6. We recommend ensuring at least 5GB of available disk space for smooth operation. No prior programming or technical expertise is necessary, as the CLI tool is ready to use. Replicating the complete experiments on the provided example dataset is as simple as executing a single command (after image build) highlighted in the README.md file. We strongly recommend building and using the Docker image. Nevertheless, it is also possible to install the \textsc{FlakeRanker} CLI tool as a Python package\footnote{\url{https://pypi.org/project/flakeranker}}. In this case, the only requirement is a minimum Python version of 3.10.

\section{Conclusion and Future Work}

This artifact \cite{aidasso_diagnosis_2025} includes numerous research results to support the findings in our paper and enable replication. It notably comprises the \textsc{FlakeRanker} CLI tool that we developed to facilitate the reuse of our RFM-based methodology for prioritizing flaky job failure categories. This CLI tool also contains a labeler module that researchers can easily reuse to obtain labeled data for future studies on the diagnosis of flaky job failures. Finally, this artifact enables extensions of our study to other contexts that face different flakiness challenges.

\section{Acknowledgment}

We acknowledge the support of Mitacs and the Natural Sciences and Engineering Research Council of Canada (NSERC).
\bibliographystyle{IEEEtran}
\bibliography{references}

\end{document}